\documentclass[aps,prl,twocolumn,showpacs,preprintnumbers,amsmath,amssymb,
floatfix,superscriptaddress]{revtex4-1}
\usepackage[english]{babel}
\usepackage{graphics}
\usepackage{graphicx}
\usepackage{comment}
\usepackage{amsmath}
\usepackage{setspace}
\usepackage{epstopdf}
\usepackage{hyperref}
\usepackage{bm}
\usepackage{xcolor}
\usepackage{SIunits}
\usepackage{booktabs}
\usepackage{tabularx}
\usepackage{array}
\usepackage{multirow}
\usepackage{braket}

\AtBeginDocument{\usepackage{booktabs}}
\newcommand{\ionm}[2]{${}^{#1}$#2${}^+$}
\newcommand{\ion}[1]{#1${}^+$}

\hypersetup{%
   pdfpagemode=None,
   pdfstartpage=1,
   pdfmenubar=true,
   pdftoolbar=true,
   colorlinks = true,
   linkcolor=blue,
   citecolor=blue,
   urlcolor=blue,
   bookmarksopen=false
 }

\newcommand{\affA}{Physikalisch-Technische Bundesanstalt, Bundesallee 100, 38116
Braunschweig, Germany}
\newcommand{\affB}{Institut f\"ur Quantenoptik, Leibniz
Universit\"at Hannover, Welfengarten 1, 30167 Hannover, Germany}

\begin{document}

\title{Coherent excitation of the highly forbidden electric octupole
transition in \ionm{172}{Yb}}

\author{H.~A.~F\"urst}\affiliation{\affA}\affiliation{\affB}
\author{C.-H.~Yeh}\affiliation{\affA}
\author{D.~Kalincev}\affiliation{\affA}
\author{A.~P.~Kulosa}\affiliation{\affA}
\author{L.~S.~Dreissen}\affiliation{\affA}
\author{R.~Lange}\affiliation{\affA}
\author{E.~Benkler}\affiliation{\affA}
\author{N.~Huntemann}\affiliation{\affA}
\author{E.~Peik}\affiliation{\affA}
\author{T.~E.~Mehlst\"aubler}\affiliation{\affA}\affiliation{\affB}\email{tanja.mehlstaeubler@ptb.de}

\date{\today}

\begin{abstract}
We report on the first coherent excitation of the highly forbidden $^2S_{1/2}
  \rightarrow{}^2F_{7/2}$ electric octupole (E3) transition in a single trapped
  $^{172}$Yb$^+$ ion, an isotope without nuclear spin. Using the transition in \ionm{171}{Yb}  as a reference, we determine the transition
  frequency to be
  $642\,116\,784\,950\,887.6(2.4)\,$Hz. We map out the magnetic field
  environment using the forbidden $^2S_{1/2} \rightarrow{}^2D_{5/2}$ electric
  quadrupole (E2) transition and determine its frequency to be
  $729\,476\,867\,027\,206.8(4.4)\,$Hz. Our results are a factor of $1\times
  10^5$ ($3\times10^{5}$) more accurate for the E2 (E3) transition
  compared to previous measurements. The results open up the way to
  search for new physics via precise isotope shift measurements and improved tests of
  local Lorentz invariance using the metastable $^2F_{7/2}$ state of \ion{Yb}.
\end{abstract}

\maketitle
\paragraph{Introduction}
The Standard Model of particle physics (SM) successfully describes many phenomena of modern physics. However, it cannot be a complete description of nature as it lacks to explain experimental evidence of, e.g., dark matter and the matter-antimatter asymmetry within the universe. Moreover, gravitation, as covered by the well-tested theory of general relativity, can up until now not be included in the SM in a renormalizable way.
Thus, tests of fundamental physics have become an important interdisciplinary field to gather new insights.
Here, table-top low energy atomic physics experiments profit from high
precision spectroscopy to make them competitive compared to high energy experiments in the search for new physics~\cite{Safronova:2018}. In particular the \ion{Yb} ion is an
excellent candidate for this, as it features a directly accessible electric octupole (E3)
transition with nHz linewidth to the electronic $F$-state, enabling the study of
violation of local Lorentz invariance (LLI) with the highest sensitivity among
accessible trapped ion
systems~\cite{dzuba2016strongly,sanner2019optical,Shaniv:2018}, similar to Tm in
neutral atom systems~\cite{golovizin2019inner}. In addition, the \ion{Yb} ion
allows for straightforward laser cooling and has two narrow electric quadrupole (E2)
transitions that can be accessed from the electronic ground state.

Access to seven stable isotopes of \ion{Yb} enables the search for new physics
via the measurement of isotope shifts (IS) of the three narrow optical
transitions and their analysis in King-plots. These measurements are
especially sensitive for the search of a possible neutron number dependent $5^\text{th}$ force, mediated
by an unknown boson, coupling electrons with neutrons~\cite{Delauny:2017,Flambaum:2018,Berengut:2018,berengut2020generalized}. Very recent measurements of the E2 transitions in \ion{Ca} with
accuracies on the order of 10\,Hz~\cite{Knollmann:2019,solaro:2020} have not yet
led to significant signatures. Here, the \ion{Yb} ion is more suitable due to
its higher sensitivity of a factor of ten~\cite{Berengut:2018}. In fact, a deviation of 3$\sigma$ from an expected
linear behavior of the King-plot of the two E2 transitions ${}^2S_{1/2}
\rightarrow {}^2D_{(3/2,5/2)}$ in even isotopes of \ion{Yb} was
found recently~\cite{Counts:2020}. The reported uncertainties of $\sim 300$\,Hz
are not sufficient yet to attribute the deviation clearly to new physics, in contrast to higher-order contributions from nuclear structure~\cite{allehabi2020using}. More
accurate measurements with uncertainties on the Hz to mHz level, comparing the IS of the E2 and E3
transitions in \ion{Yb} will provide a higher sensitivity to new physics as
they are of different electronic type~\cite{Berengut:2018} and the uncertainties of the
isotope mass can be eliminated using three
transitions~\cite{berengut2020generalized}. However, the required precision in the Hz range has not been realized so far in even isotopes of \ion{Yb}, as they are magnetic field sensitive to $1^{\text{st}}$ order. With this work, we open up the even isotope \ionm{172}{Yb} for clock spectroscopy and demonstrate frequency uncertainties at the Hz level for both the ${}^2S_{1/2}
\rightarrow {}^2D_{5/2}$ (E2) and the ${}^2S_{1/2}
\rightarrow {}^2F_{7/2}$ (E3) transition and by this improving the literature values by at least five orders of magnitude~\cite{Roberts:1997,Taylor:1997}.

The first laser excitation of the highly forbidden E3 transition in
\ionm{172}{Yb} ions was carried at the National Physical Laboratory (NPL) and
led to an uncertainty of 0.7\,MHz in the transition frequency and an excitation
rate of about 0.03\,s$^{-1}$ on resonance~\cite{Roberts:1997}. Here,
we demonstrate the first coherent quantum state control, by achieving
90\,\% excitation probability for Rabi spectroscopy with 42\,ms long
pulses. The coherence time is found to be 190(27)\,ms and the achieved minimum
linewidth of the transition is 6.0(6)\,Hz. We measure the frequency difference
to the E3 clock transition in \ionm{171}{Yb} ($F=0\rightarrow3$, $\Delta m_F=0$)
with an uncertainty of 2.3\,Hz and derive the transition frequency with an
uncertainty of 2.4\,Hz. For the
E2 transition, we obtain an uncertainty of 4.4\,Hz. For both frequency measurements we discuss the uncertainties and show that they can be kept at sub-Hz level for transition frequency differences between different isotopes. Our results will open up a way to precise IS
measurements in \ion{Yb} and for a rapid and defined preparation of the
$F_{7/2}$ state for sensitive tests of LLI, increasing current limits by two
orders of magnitude~\cite{Shaniv:2018,sanner2019optical}.

\begin{figure}
	\includegraphics[width=0.45\textwidth]{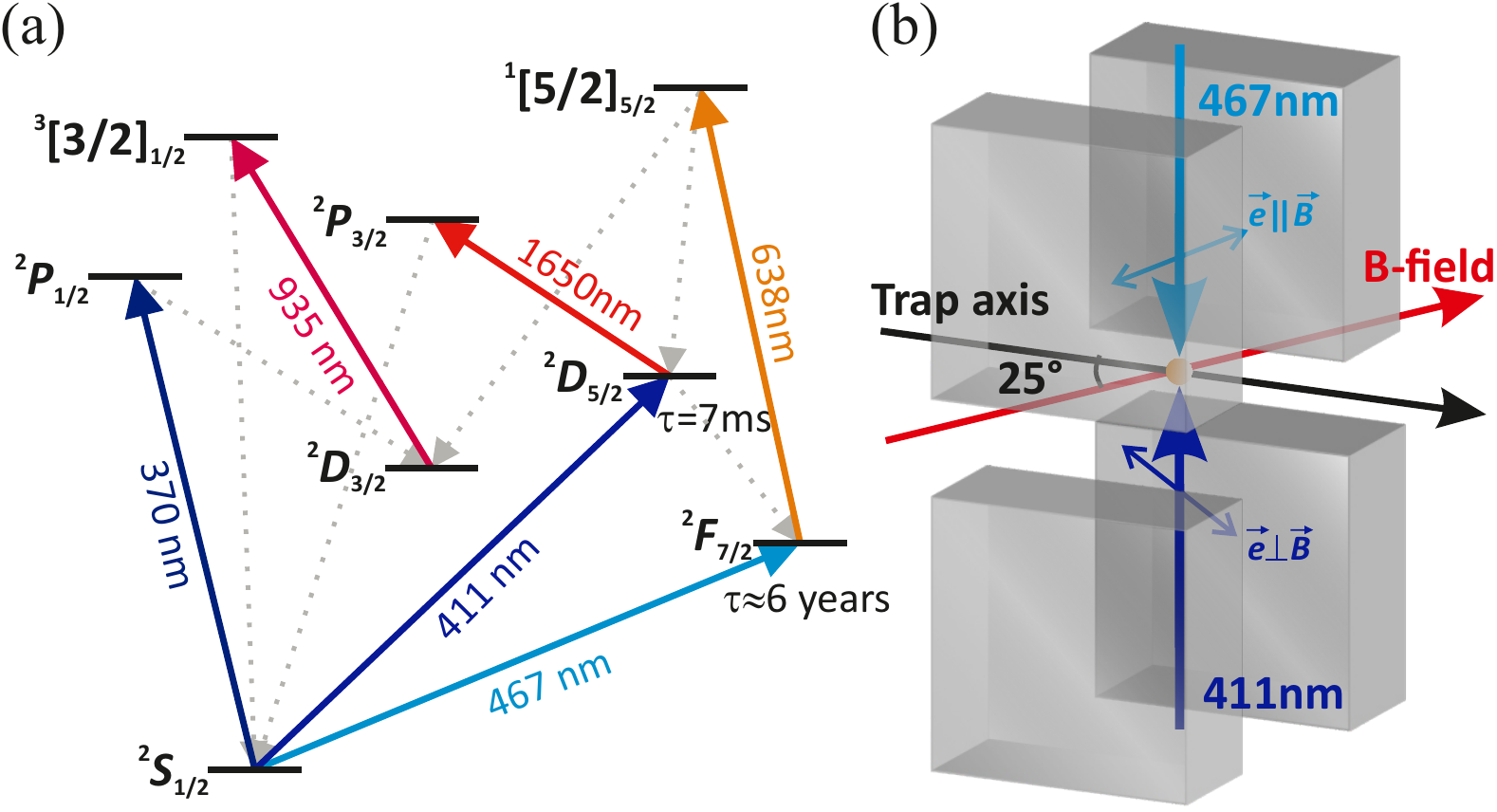}
    \caption{(a) Reduced level scheme of \ionm{172}{Yb}. Doppler-cooling and
    repumping is carried out on the transitions near $370\,\nano\meter$ and
    935\,nm, respectively. Optical pumping into the $m_{J}$ electronic ground
    states is done via a $\sigma$-polarized $370\,$nm beam. Excitation of the
    metastable $D_{5/2}$ and $F_{7/2}$ states via the transition near 411\,nm
    and 467\,nm are detected using fluorescence imaging on the cooling
    transition, followed by repumping using the transitions near 1650\,nm and
    638\,nm. (b) Laser access and $\vec{B}$-field orientation for the probe beams of polarization $\vec{e}$
    within the ion trap segment (gray).}
	\label{fig:scheme}
\end{figure}

\paragraph{Experimental setup}
Our experiment is carried out in a radiofrequency (rf) Paul trap, as described
in Ref.~\cite{Keller:2019}. Single ions are Doppler-cooled to 0.5(1)\,mK on the
transition near $370\,\nano\meter$, assisted by a repumper laser near 935\,nm
(see Fig.~\ref{fig:scheme}). For the interrogation of the E2 transition, we use
a frequency-doubled diode laser near $822\,\nano\meter$, locked to a cavity with
a fractional instability of $5\times 10^{-16}$ at $10\,\second$ averaging
time~\cite{Keller:2014}, providing the short-term stability of the system. The
light is amplified with an injection-locked laser diode and frequency doubled in
a periodically poled potassium titanyl phosphate (PPKTP) crystal to
$411\,\nano\meter$. A maximum power of about 0.6\,mW is focused down to a waist of
$83\,\micro\meter$ at the position of the ion.

Coherent excitation of the E3 transition requires an ultra-stable, high
intensity laser source near 467\,nm.  For that we use a seed laser power of
about 0.5\,mW near 934\,nm from the probe laser of the \ionm{171}{Yb} single ion
optical clock~\cite{Huntemann:2016} via a stabilized fiber link. We use an
acousto-optical modulator (AOM) near 2.3\,GHz to bridge the frequency difference
and two injection-locked laser diodes for light amplification.  The light is
frequency-doubled in a periodically poled LiNbO$_3$ (PPLN) waveguide to 467\,nm.
We obtain about 8\,mW of probe light with beam waists of $(w_x,
w_y)=(26(3),38(3))\,\micro\meter$ at the ion. During the spectroscopic
interrogation, the 934\,nm laser is referenced to the E3 clock transition of
\ionm{171}{Yb}, a recommended secondary representation of the SI
second~\cite{Huntemann:2012,Riehle:2018}.

For both probe lasers, power stabilization, switching and frequency
tuning is performed via AOMs. Spectroscopy is
carried out after optical pumping using circularly polarized cooling light to
prepare population in one of the $m_J=\pm1/2$ electronic ground states,
followed by the respective probe laser pulse.
Excitation is detected by the absence of fluorescence at the 370\,nm cooling
transition (electron shelving). Repumping is carried out as shown in Fig.~\ref{fig:scheme}.

\paragraph{Results on the $^2S_{1/2} \rightarrow{}^2D_{5/2}$ transition}
By coherent control of the E2
transition near 411\,nm we achieve an excitation of over
95\,\% for a $\pi$-pulse of $t_\pi=8.6\,\micro$s duration.
Using this transition
we map out and minimize magnetic field noise at the ion position, by actively stabilizing the magnetic field
as described in~\cite{Suppl:2020}.
Further, we align the focus of the 467\,nm
spectroscopy beam onto the ion by maximizing the induced ac Stark shift on the
E2 ($m_J=-1/2\rightarrow-5/2$) transition. A frequency shift of 2.1(1)\,kHz at a peak intensity
of $I_{467}=5.2(7)\times 10^6\,$W/m$^2$ is observed for a polarization of the 467\,nm beam parallel
to the quantization axis (see Fig.~\ref{fig:scheme}(b)).

We obtain spectra of the $m_J = \pm1/2 \rightarrow \pm 5/2$ transitions by measuring population in the $D_{5/2}$ as a function of the frequency of the excitation laser. Alternating of the transitions cancels out slow
drifts in the magnetic field. We use excitation pulsed of $t_\pi=1.6\,$ms leading to about 80\,\% excitation probability.
Averaging 40 spectra measured within 1.5\,h leads to a statistical uncertainty of 2.5\,Hz. During this period the 822\,nm master laser frequency is calibrated using the primary frequency standard CSF2~\cite{wey18} via an
optical frequency comb~\cite{Falke:2011,schwarz2020long} using a hydrogen maser as a flywheel oscillator. A drift of the 822\,nm locking cavity was accounted for via the data obtained by the frequency comb measurement. The short averaging time leads to an additional statistical uncertainty of 1.3\,Hz. We determine the center frequency of
the E2 transition to be $729\,476\,867\,027\,206.8(4.4)\,$Hz,
improving upon the uncertainty given in literature by a factor of
$1.0\times10^5$~\cite{Taylor:1997}. The frequency is corrected for known systematic shifts as discussed at the
end of the letter. 

\begin{figure}[t]
	\centering
	\includegraphics[width=0.95\columnwidth]{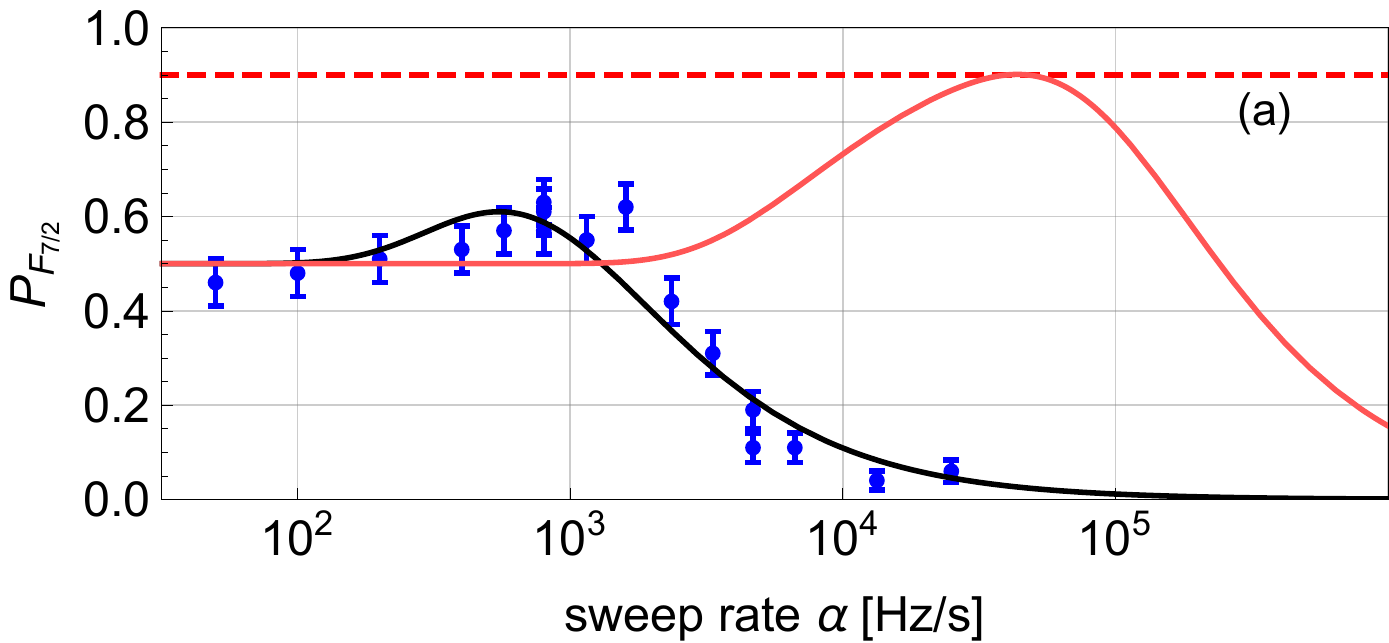}
	\includegraphics[width=0.95\columnwidth]{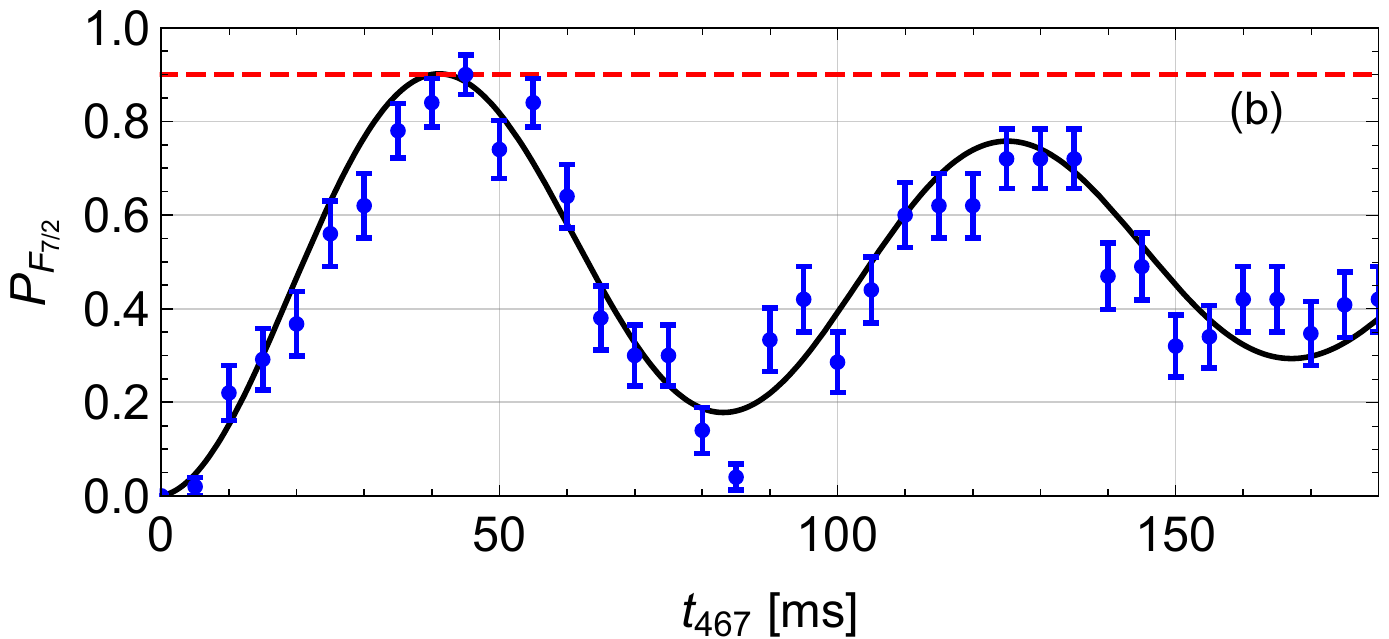}
  \caption{(a) Excitation probability for a RAP pulse covering the resonance
   within a window of 200\,Hz vs.\ sweep rate $\alpha$ (blue) and a fit
   according to Eq.~\ref{eqn:rapmodel} (black). (b) Rabi flop on the
   $|S,-1/2\rangle\rightarrow|F,-1/2\rangle$ E3 transition (blue) along with
   a sinusoidal fit with exponential envelope (black), leading to a Rabi
   frequency of $11.90(14)\,$Hz and a decoherence time of $\tau$ = 190(27)\,ms.
   Each point was averaged over 50 measurements. In both plots the red dashed
   line indicates $90\,\%$ excitation probability. This required a Rabi
   frequency of 130\,Hz when using a RAP instead ((a), solid red).}
   \label{fig:flop}
\end{figure}

\paragraph{Excitation of the $^2S_{1/2} \rightarrow{}^2F_{7/2}$
transition}
To initially observe the E3 transition within the large uncertainty interval of
1.4\,MHz~\cite{Roberts:1997}, we employ
a RAP technique. The technique allows for
a robust excitation of a transition in the presence of noise,
e.g., from the laser source or magnetic fields~\cite{Noel:2012}. This is achieved by sweeping the
laser frequency adiabatically across the resonance during the spectroscopy
pulse. For a reasonably slow sweep rate $\alpha \ll \Omega \cdot \Gamma$ (with $\Gamma < \Omega$), the transition
probability reaches at least 50\,\%, without exact knowledge of the present noise
figure $\Gamma$ and origin (e.g. frequency fluctuations of the probe laser) and exact resonant Rabi frequency $\Omega$~\cite{Noel:2012,Lacour:2007}. We choose the $\Delta m_J=0$ transitions as they
are least magnetic field sensitive ($\pm6\,$kHz/$\micro$T) in a field of
$6.5\,\micro$T. After minimization
of environmental noise sources, we use pulses of maximum intensity and sweep the
detuning across the expected resonance in windows of $\Delta f_\text{RAP}
= 200\,$Hz for a pulse length of $t_\text{RAP}=1\,$s. We optimize the excitation
probability $P_{F_{7/2}}$ by varying
the sweep rate $\alpha = \Delta f_\text{RAP}/t_\text{RAP}$, as shown in
Fig.~\ref{fig:flop}(a) (blue). The data is fitted (black) using the model in
Refs.~\cite{Noel:2012,Lacour:2007},
\begin{align}
P_{F_{7/2}}=\left(1-e^{-\frac{\Omega ^2}{4 \alpha }}\right) e^{-\frac{\Gamma \Omega }{2 \alpha }}+\frac{1}{2} \left(1-e^{-\frac{\Gamma \Omega }{2 \alpha }}\right)\,.
\label{eqn:rapmodel}
\end{align}
The fit leads to $\Omega/(2\pi) = 9.6(5)$\,Hz and a noise figure of
$\Gamma/(2\pi)=3.0(9)$\,Hz.
At this Rabi frequency a maximum excitation probability of $P_{F_{7/2}} = 60\,\%$ is obtained for pulse times of 360\,ms ($\alpha = 556\,$Hz/s), limited by the noise figure $\Gamma$ being similar to $\Omega$.
However, the RAP technique can serve as a helpful tool to efficiently find the
transition in other isotopes
for the search of new physics when looking for
anomalies in IS~\cite{Berengut:2018,Counts:2020}.

To achieve a higher spectroscopic resolution and faster population transfer,
as required for an efficient test of LLI in the $F$-manifold of \ion{Yb}, we use
Rabi spectroscopy with pulses of constant frequency and intensity. Within
the 200\,Hz window identified with RAP, the resonance can be found easily. On
resonance, Rabi spectroscopy at maximum power leads to $P_{F_{7/2}}=90.0(1)\,\%$, as depicted in Fig.~\ref{fig:flop}~(b) (blue), where Rabi-oscillations of the $F$-state population are shown along with a sinusoidal
fit (black) with exponential decaying envelope. The fit leads to decoherence time of
$\tau=190(27)\,$ms and $\Omega/(2\pi)=11.90(14)\,$Hz, similar to the RAP
model. Note that achieving a similar excitation probability using the RAP method
would require an intensity of a factor 180(20) higher (Fig.~\ref{fig:flop}(a),
solid red) to satisfy $\Omega \gg \Gamma$.

\begin{figure}[htbp]
	\centering
	\includegraphics[width=0.95\columnwidth]{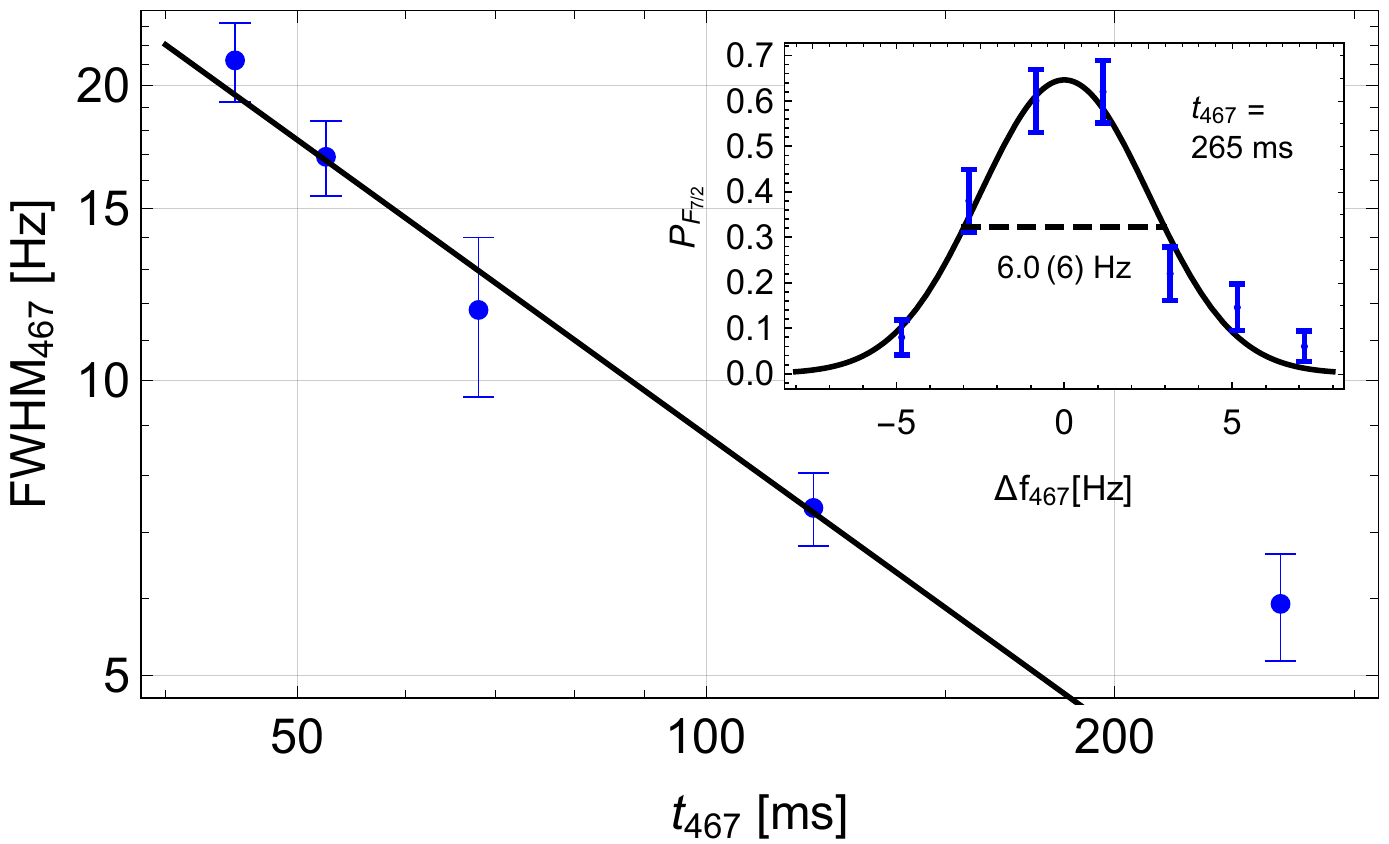}
	\caption{Observed transition linewidth FWHM${}_{467}$ versus length $t_{467}$ of
the spectroscopy pulse (blue) along with
the Fourier limited linewidth for the used pulseshape (black). The inset shows the
spectrum at $t_{467}=265\,$ms (blue) along with a Gaussian fit
(black).}
	\label{fig:fourierlimit}
\end{figure}

To investigate the minimum achievable linewidth of the transition, we varied the intensity of the spectroscopy beam and the resonant $\pi$-pulse time accordingly. The fitted
linewidths of the spectra (FWHM: full width at half maximum) for several pulse
times are shown in Fig.~\ref{fig:fourierlimit} (blue).
The black line corresponds to the Fourier
limit of $\Delta f_{\pi}\approx0.89/t_{\pi}$ for rectangular shaped
$\pi$-pulses. The inset shows a spectrum for $t_{467}= 265\,$ms, deviating from
the Fourier limit by additional 2.5(7)\,Hz due to incoherent contributions of the magnetic field
during the 180\,s of measurement time. The additional broadening is in agreement
with the noise level extracted from the RAP measurements.

To determine the frequency, we alternately measure the spectra of the
two $\Delta m_J=0$ Zeeman components using $\pi$-pulses of 124\,ms to achieve $1^\text{st}$ order magnetic field
insensitivity. We average the data over 16 spectra measured within 1\,h, leading to a statistical uncertainty of 0.13\,Hz.
We determine the frequency shift to the
$F=0\rightarrow F=3$ ($\Delta m_F = 0$) clock transition to be
$f_{E3}^{(172)}-f_{E3}^{(171)}= -4\,711\,821\,757.3(2.3)$\,Hz.

For the determination of the transition frequency, the uncertainty of the recommended value of the reference frequency in \ionm{171}{Yb} of 0.39\,Hz~\cite{Riehle:2018} is taken into account. This yields $f_{E3}^{(172)} = 642\,116\,784\,950\,887.6(2.4)\,$Hz. The frequencies are corrected for known systematic shifts as discussed in the following.

\paragraph{Systematic shifts and uncertainties}
Major contributions to the systematic shifts and their uncertainties are summarized in
Tab.~\ref{tab:errors}.

\begin{table}[]
	\centering
	\caption{Leading systematic frequency shifts $\delta\nu$ and related uncertainties $\mu$ in Hz
for the $^2S_{1/2}\rightarrow{}^2D_{5/2}$ quadrupole (E2) and
$^2S_{1/2}\rightarrow{}^2F_{7/2}$ octupole (E3) transition in \ionm{172}{Yb}.}
	\begin{tabular}{l>{\centering}p{0.055\textwidth}>{\centering}p{0.055\textwidth}>{\centering}p{0.055\textwidth}>{\centering\arraybackslash}p{0.055\textwidth}}
		\hline
		\hline
		\multicolumn{1}{c}{} & \multicolumn{2}{c}{E2} & \multicolumn{2}{c}{E3} \\ \hline
		Effect & $\delta\nu$ & $\mu$ & $\delta\nu$ & $\mu$\\ \hline
		935\,nm ac Stark & 8.8	& 2.7 & \multicolumn{2}{c}{--} \\
		quadrupole shift & -9.9 & 2.0 & \multicolumn{1}{c}{-0.07} & 0.01  \\
		Stark shift (probe light) & 0.003 & 0.002 & \multicolumn{1}{c}{33.0}  &   2.3 \\ 
		blackbody radiation & -0.24 & 0.11 & \multicolumn{1}{c}{-0.07} &0.03  \\
		total & -1.3 & 3.4 & \multicolumn{1}{c}{32.9} & 2.3 \\ \hline\hline
	\end{tabular}
	\label{tab:errors}
\end{table}
For both transitions, an electric quadrupole shift, resulting from the dc
trapping potential~\cite{Itano:2000} and stray electric fields has to be
considered. The shift is calculated to be $-9.9(2.0)\,$Hz for the E2 and
$-0.07(1)\,$Hz for the (E3) transition for an axial trap frequency of
213(5)\,kHz and an angle of $25(5)^{\circ}$ between the trap axis and the
magnetic field. The quadrupole moments are taken from
Refs.~\cite{Nandy:2014,Lange:2020}, respectively.  The effect of stray electric
fields can be estimated from the dc fields required to compensate excess
micromotion and is at least an order of magnitude smaller and is included in the
uncertainty budget.  Note that our trap frequencies, and thus field gradients,
show a long term stability of below $10^{-4}$ and quadrupole shifts can be
measured and monitored at the mHz level if needed~\cite{Dube:2005}.

The large intensity of the probe light leads to a significant ac Stark shift of
the E3 transition. We deduce the shift by measuring the resonance frequency at
different optical powers and extrapolate to zero as described
in~\cite{Suppl:2020}. With this we obtain an ac Stark shift of 33.0(2.3)\,Hz for
the 124\,ms $\pi$-pulses used in the frequency measurement. For the E2
transition, the ac Stark shift is much smaller. Using the information provided
by Refs.~\cite{feldker:2017,Roy:2017,Schneider:2005,Roy:2020}, we compute the
differential polarizability of the used $m_J = \pm1/2\rightarrow \pm5/2$
transition at 411\,nm to be $2.8(1.5)\times10^{-39}\,\text{Jm}^2\text{V}^{-2}$,
leading to a shift of 0.003(2)\,Hz for the used $\pi$-pulses of 1.6\,ms.

An additional Stark shift stems from blackbody radiation (BBR)~\cite{Keller:20191}. At an effective temperature at the ion position of
297(3)\,K~\cite{Dolevzal:2015,Nordmann:2020}, the BBR shift is calculated to be -0.24(11)\,Hz for the E2 and $-0.07(3)$\,Hz for the E3 transition, limited by the instability of the lab
temperature. The involved differential polarizability for the E2
transition is computed to be $-4.5(2.3)\times10^{-40}\,\text{Jm}^2\text{V}^{-2}$
using Refs.~\cite{feldker:2017,Roy:2017,Schneider:2005,Roy:2020}, where the static value of $-4.3(2.2)\times10^{-40}\,\text{Jm}^2\text{V}^{-2}$ was corrected for the BBR spectrum~\cite{Farley:1981,Schneider:2005}, whereas for
the E3 transition the value can be found in Ref.~\cite{Huntemann:2012}.
Due to the lack of a fast laser shutter during the frequency measurement, the
935\,nm repumper beam was present during the E2 interrogation, leading to an ac
Stark shift. We determine the shift to be 8.8(2.7)\,Hz in a separate measurement
with interleaved interrogation of the E2 transition with and without the 935\,nm
laser field.
Additional systematic shifts $<0.02$\,Hz are discussed
in the supplemental material~\cite{Suppl:2020}.

Combining statistical and systematic uncertainties we obtain a total uncertainty
of 4.4\,Hz and 2.4\,Hz for the frequencies of the E2 and E3 transitions,
respectively.

\paragraph{Conclusion}
We report on the first coherent excitation of the highly forbidden E3 transition
in the even isotope \ionm{172}{Yb} with an excitation probability of 90\,\%
within 42\,ms. Fast and reliable state preparation in the $F_{7/2}$ state Zeeman
manifold is an important requirement for an improved test of LLI with
well-controllable trapped ion Coulomb crystals of
\ionm{172}{Yb}~\cite{Keller:20191,Shaniv:2018} to enable a high duty-cycle and
give a high signal-to-noise ratio for the LLI signal.

We measured the frequencies of the E2 transition near 411\,nm and the E3
transition with an uncertainty of 4.4\,Hz and 2.4\,Hz, respectively. In
particular, in Ref.~\cite{Berengut:2018} it was proposed that reaching the Hz
level in uncertainties of the isotope shifts (IS) of these transitions should
allow one to investigate the so-called ${}^8$Be
anomaly~\cite{Krasznahorkay:2016}.

Probing the sub-Hz regime in IS for the E3 and both E2 transitions is predicted
to provide further insights, as the measurement of a third transition will
eliminate mass uncertainties and higher-order Standard Model contributions,
which can also lead to
a nonlinearity~\cite{berengut2020generalized,allehabi2020using}.
Sub-Hz accuracy of the systematic shifts can be achieved by transferring
techniques demonstrated in the \ionm{171}{Yb} clock
spectroscopy~\cite{Tamm:2014,Huntemann:2016} to the
even isotopes.
Alternating interrogation of different
isotopes suppresses common mode effects~\cite{Counts:2020} such as the
quadrupole shift, BBR shift and trap-rf-related Stark shift, as the large uncertainties of the quadrupole moments and of the polarizabilities drop out. The BBR shifts can be kept reproducible at the mHz level if the temperature is stabilized to $\Delta T < 0.3\,$K. For the quardupole shift, the influence of the uncertainties of electric field gradient and angle of quantization axis can be kept stable at sub-mHz levels~\cite{Dube:2005}. Alternatively,
entanglement of different co-trapped isotopes as presented in
Ref.~\cite{Manovitz:2019} can be applied to \ion{Yb} and has proven to reach mHz
accuracies in IS measurements of~\ion{Sr}.

\begin{acknowledgments}
We kindly acknowledge help from Stefan Weyers in the frequency measurement of
  the quadrupole transition and Atish Roy for support with computation of the
  $D_{5/2}$ state polarizabilities. We thank Michel Wolf and Tjeerd J. Pinkert
  for work on the magnetic field stabilization and Michael Drewsen for fruitful
  discussions. This project has been supported by the Deutsche
  Forschungsgemeinschaft (DFG, German Research Foundation) through grant CRC SFB
  1227 (DQ-mat, project B03) and through Germany's Excellence Strategy
  – EXC-2123 QuantumFrontiers - 390837967. This work has been supported by the
  EMPIR project 18SIB05 ``Robust Optical Clocks for International Timescales''.
  This  project has received funding from the EMPIR programme co-financed by the
  Participating States and from the European Unions Horizon 2020 research and
  innovation programme. This work has been supported by the
  Max-Planck-RIKEN-PTB-Center for Time, Constants and Fundamental Symmetries. 
\end{acknowledgments}

\section{Supplemental material}

\subsection{Magnetic field noise cancellation}
Excitation of a nanohertz-wide magnetic field sensitive transition requires
a high degree of control over the magnetic field environment. In our experiment
we reduce fluctuations of the magnetic field by employing a feedback loop with
a bandwidth of about 10\,Hz. The field is measured with a sensor nearby the
vacuum chamber. The loop provides feedback via coils in three dimensions. The
arrangement is able to reduce typical quasistatic magnetic field variations to
below $4.7(5)\,$nT, limited by the distance of about 5.5\,cm between the sensor
and the ion. During the spectroscopic measurements we ensured that nearby
transient disturbances were avoided, limiting the variations to less than
$0.4\,$nT.

\begin{figure}[htbp]
	\includegraphics[width=0.45\textwidth]{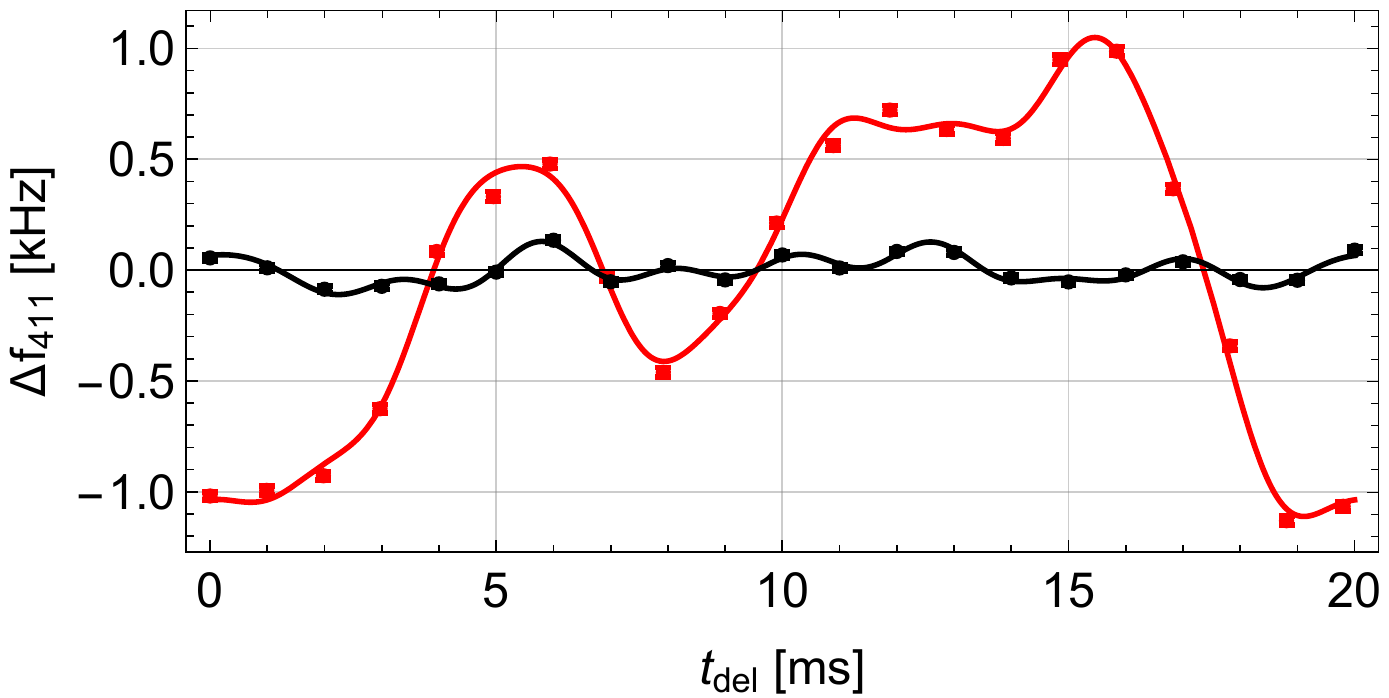}
  \caption{Peak positions of the employed E2
  transition vs.\ delay time of the spectroscopy pulse
  from the line trigger. The stray magnetic field from the ac line noise shifts
  the transition by about 2\,kHz peak to peak during a line cycle (red),
  corresponding to $70\,$nT. Using a feed-forward approach this is reduced
  by a factor of 5 (black). Error bars stem from fitting errors of the peak
  position. The data is shown along with a sinusoidal fit, taking into account
  the first nine harmonics of the 50\,Hz.}
	\label{fig:dfvsdelay}
\end{figure}

The ac mains voltage at 50\,Hz and its higher harmonics lead to a periodic
distortion of about $70\,$nT peak-to-peak (pp) on top of the quantization
magnetic field at the ion. This can lead to sidebands and reduced excitation
probability on the narrow E3 transition. Employing the
$^2S_{1/2}\rightarrow{}^2D_{5/2}$ E2 transition, we sample this contribution by
measuring the resonance frequency of the $m_J=-1/2\rightarrow -5/2$ Zeeman
component as a function of a delay relative to a line trigger, as shown in
Fig.~\ref{fig:dfvsdelay} (red). To compensate for this shift, we employ a shunt
circuit that is fed by a waveform, derived by a microcontroller, similar to the
approach in Ref.~\cite{merkel2019magnetic}.
The circuit subtracts a
time-dependent current from the original driving current in the quantization
field coils leading to a suppression of more than a factor of 5, see
Fig.~\ref{fig:dfvsdelay} (black). From simulations we find this suppression to
be sufficient for excitation of the $\Delta m = 0$ E3 transitions without significant
sideband amplitudes. The method is limited by the day-to-day fluctuations of
the ac line noise within the building. Further incoherent contributions to the
magnetic field noise are measured to be $\leq 0.4\,$nT (pp).

\subsection{Extrapolation of the ac Stark shift of the E3 transition}
\begin{figure}[ht]
	\includegraphics[width=0.45\textwidth]{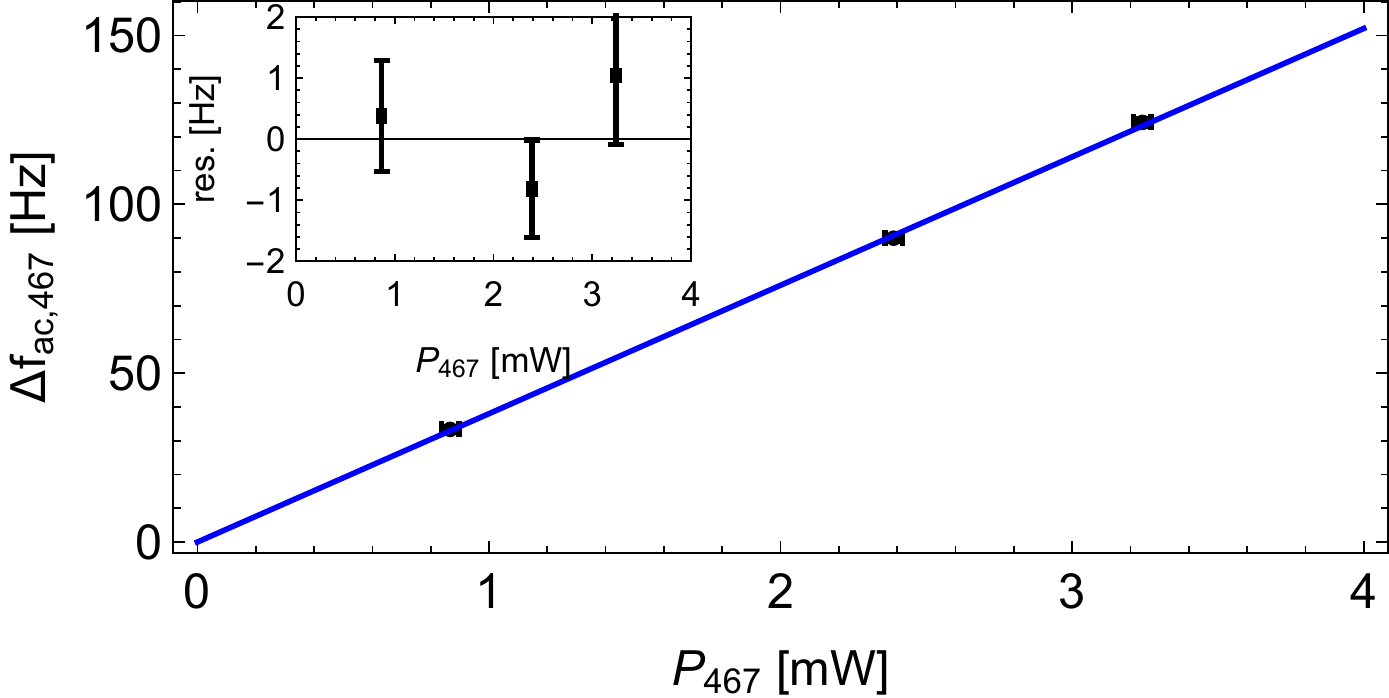}
  \caption{Light-induced AC-Stark shift of the E3 transition vs.\ laser power
  proportional to the intensity at the ion position (blue) along with a linear
  fit (black).}
	\label{fig:slope}
\end{figure}

To measure the light-induced ac Stark shift of the E3 transition, we
measure the resonance frequency of the transition for different beam powers $P_{467}$. Assuming
a linear behavior~\cite{Huntemann:2012}, we fit the slope to extrapolate to
zero intensity. Therefore we
determine the beam powers using a calibrated photo-diode. Calibration errors are
negligible on the timescales of the measurement. The error in the power reading
stems from detection noise and is less than 0.03\,mW whereas the linearity of the photo-diode in the used range can be neglected. Errors in the frequency
determination stem from fit errors of the respective excitation spectra. Results are shown in
Fig.~\ref{fig:slope} (black) along with the linear fit (blue). The inset shows
the fitting residuals.
The fit leads to a shift of $\Delta f_{\text{ac},467}
= 38.0(8)\,\text{Hz/mW}\times P_{467}
+ 0.0(1.9)\,\text{Hz}$ for the beam parameters used in the
experiment, corresponding to $\Delta
f_{\text{ac},467}=0.66(6)$\,Hz$^{-1} \Delta f_{\text{Fourier}}^2$, with $\Delta
f_\text{Fourier}$ the Fourier limited linewidth (FWHM) of the spectroscopy
$\pi$-pulse, in agreement with the value for
\ionm{171}{Yb}~\cite{Huntemann:2012}. Drifts due to beam pointing were measured to be $\leq0.4$\,Hz during the frequency measurements and are included in the computation of the total uncertainty of the ac Stark shift.

\subsection{Additional frequency shifts of the transitions}
The major contributions of the systematic frequency shifts and their
uncertainties are discussed in the
main text. Contributions that are negligible on the current measurement accuracy
are given here for completeness. In addition, the gravitational redshift that stems from comparison of the probe laser frequencies to the frequency references is given.
\paragraph{Second-order Doppler shift}
To compute the $2^\text{nd}$ order Doppler shift~\cite{Keller:20191}, we measure
the ion temperature to be 0.5(1)\,mK via carrier Rabi thermometry on the
411\,nm quadrupole transition. Additionally,
excess micromotion of the ion contributes to the shift. We compensate this
micromotion to below $E_\text{rf}=(50(20),60(10),125(25))$\,V/m of RF electric
field amplitude at the position of the ion. Thus we compute the absolute value of the
$2^\text{nd}$ order Doppler shift to be $\leq 0.002$\,Hz for both transitions.

\paragraph{Second-order Zeeman shift}
For both transitions, we average over two Zeeman components to be $1^\text{st}$
order Zeeman insensitive. The $2^\text{nd}$ order Zeeman shifts in the even
isotopes are negligible due to the lack of hyperfine structure and large fine
structure splittings $\Delta_\text{FS}$ of the $D$ and $F$
states~\cite{Herschbach_Linear_2012,NIST}. The shift for alkali-like ions is
given by $\Delta f_{\text{Zeem},\text{2nd}} = 2 \mu_\text{B}^2/(3 h^2
\Delta_\text{FS})B^2$~\cite{Herschbach_Linear_2012}, where $\mu_B$ is the Bohr
magneton, $h$ the Planck constant and $B=6.5\,\micro$T the absolute value of the
magnetic field. For the splittings between the $D$-states and the $F$-states,
these shifts are $<0.02$ and $<0.002\,$Hz, including ac contributions due to
micromotion and currents of the trap drive~\cite{Gan:2018}.

\paragraph{Second-order Stark shift due to trap rf}
The $2^\text{nd}$ order Stark shift caused by the electric fields of to the trap rf computes to -0.008(5)\,Hz for the E2 and -0.003(1)\,Hz for the
E3 transition, including the effect of uncompensated excess micromotion and intrinsic micromotion due to the the finite ion temperature~\cite{Keller:20191}. The involved static differential polarizability for the E2
transition is computed to be $-4.3(2.2)\times10^{-40}\,\text{Jm}^2\text{V}^{-2}$
using Refs.~\cite{feldker:2017,Roy:2017,Schneider:2005,Roy:2020}, whereas for
the E3 transition the value can be found in Ref.~\cite{Huntemann:2012}.

\paragraph{Gravitational redshift}
The values of the transition frequencies were corrected for gravitational
redshifts with uncertainties of 0.004\,Hz and 0.005\,Hz for
the E2 and E3 transition, respectively~\cite{Denker:2018}.\\

\end{document}